\documentclass[11pt]{article}

\usepackage[margin=1in]{geometry}
\usepackage[T1]{fontenc}
\usepackage[utf8]{inputenc}
\usepackage{lmodern}
\usepackage{microtype}
\usepackage{times}

\usepackage{enumitem}
\usepackage{array}
\usepackage{booktabs}
\usepackage{graphicx}
\usepackage{xcolor}

\usepackage{amsmath,amssymb,amsthm}
\usepackage{mathtools}

\usepackage[numbers]{natbib}
\usepackage[colorlinks,bookmarksopen,bookmarksnumbered,citecolor=red,urlcolor=red,linkcolor=red]{hyperref}

\usepackage{authblk}

\IfFileExists{arxiv.sty}{\usepackage{arxiv}}{}

\title{Resource-Aware Neuro-Symbolic Reasoning for Local Small Language Models}

\author[1]{Carlos Ram\'irez}
\author[1]{Abel Alvarez}
\affil[1]{Pontificia Universidad Javeriana Cali, Cali, Colombia}
\affil[ ]{\texttt{carlosovalle@javerianacali.edu.co} \quad \texttt{abel.alvarez@javerianacali.edu.co}}

\date{} 

\begin{document}
\maketitle

\begin{abstract}
Small language models can run locally on consumer hardware, but structured
reasoning often pushes users toward repeated sampling or larger models. This
article studies a bounded neuro-symbolic alternative for local inference: a
model translates a problem into typed finite-domain rules and constraints, a
symbolic layer checks traceability and consistency, and a deterministic solver
performs the reasoning step. The resulting Verifiable Formalization and Repair
pipeline (VFR-LLM) tests when symbolic verification can replace repeated
sampling without weakening accuracy. We evaluate it through LM Studio on Apple
Silicon, using Qwen3-4B-2507 as the primary model, with Phi-4-mini-reasoning and
Gemma-3n-E4B as robustness checks. On 120 generated pure-precedence problems,
Qwen VFR-LLM achieves 0.983 accuracy, versus 0.700 for serial self-consistency
using one model call instead of five. On a 120-instance BBH-derived extended
logical-deduction subset, it reaches 0.933 versus 0.283. The advantage persists
against a stronger cost-aware adaptive self-consistency baseline, which lowers
sampling cost but not the single-call accuracy gap. Gemma reproduces the same
model-dependent boundaries and Phi is negative on typed constraints. The
contribution is bounded: finite-domain logic can replace repeated local sampling
for some structured tasks, saving model calls and serial latency, with
stratum-dependent token savings.
\end{abstract}


\section{Introduction}

Small language models (SLMs) can run locally on consumer hardware, improving
privacy, offline availability, and operational cost. However, local models are
constrained by memory, latency, and limited model capacity. A common strategy
for improving reasoning accuracy is to sample several independent answers and
aggregate them. This strategy, often called self-consistency, can improve
performance on reasoning benchmarks, but it also multiplies local model calls,
generated tokens, and wall-clock time when samples are executed serially.
These constraints are a local instance of broader deployment challenges around
model scale, latency, and resource demand in contemporary language models
\cite{kumar2024large,wang2025edge}. They also connect to the emerging small
language model literature, which treats local or edge execution as a distinct
design target rather than merely a compressed version of cloud-scale inference
\cite{corradini2025slmreview,song2025smallthinker}.

This work studies a narrower alternative: replacing part of repeated local
neural generation with symbolic verification and solver-based reasoning. The
proposed system asks the SLM to translate the problem into a typed finite-domain
rule-and-constraint representation, checks whether extracted facts and
constraints are grounded in the original problem, executes the reasoning step
with a deterministic solver, and applies deterministic repair only when
diagnostics identify a local formalization failure.

The broader goal is not to make a small local model behave like a general
theorem prover. The goal is practical: identify a logical target language that
is expressive enough for structured tasks, small enough for a local model to
produce reliably, and cheap enough for a consumer machine to verify. This
matters because a resource-aware claim is easy to make vaguely: if a local
model already answers a simple problem in one short generation, adding a solver
is unlikely to be the cheapest option.

The intended comparison is therefore conditional. The proposed Verifiable
Formalization and Repair pipeline (VFR-LLM) is meant to compete with repeated
local sampling when a single answer is not trusted, when failures need to be
diagnosed, or when a user needs an auditable reasoning trace rather than only a
plausible final string. Direct one-shot answering remains a necessary baseline
and, in some strata, a winner. The empirical contribution of this paper is to
make that boundary visible rather than assuming that adding logic automatically
lowers compute.

\paragraph{Contributions.}
This paper makes four contributions. First, it defines a traceable
finite-domain rule-and-constraint representation based on typed facts, safe
Horn-style rules, and solver-facing constraints. Second, it implements a
verification and repair loop that uses source spans and solver diagnostics to
identify formalization failures before final answer generation. Third, it
reports a local evaluation matrix with Qwen3-4B,
Microsoft Phi-4-mini-reasoning, and Gemma-3n-E4B through LM Studio, including
MLX/Metal, a CPU-only Qwen typed condition, and public-source BBH-derived
subsets with pairwise and absolute-position constraints. Fourth, it reports a
mixed result that narrows the claim: VFR-LLM strongly supports a
resource-reduction claim for Qwen on pure-precedence and BBH-derived structured
tasks, gives weaker but positive evidence for Gemma on generated precedence,
and remains blocked or marginal in several typed and cross-model conditions.

\section{Research Questions and Hypotheses}

\textbf{RQ1.} Can a verifiable local SLM--symbolic solver pipeline based on a
typed finite-domain rule-and-constraint formalism reduce generated tokens,
repeated model calls, and serial latency relative to self-consistency while
preserving or improving answer accuracy?

\textbf{RQ2.} Under which classes of structured reasoning problems does the
symbolic layer improve the accuracy--cost trade-off, and under which classes
does direct one-shot answering remain preferable?

\textbf{RQ3.} What are the dominant failure modes in translating natural
language into typed facts, Horn-style rules, and finite-domain constraints?

\textbf{RQ4.} Can solver-guided diagnostics reduce the number of additional SLM
calls required to repair incorrect formalizations?

\paragraph{Hypothesis.}
For tasks with explicit constraints, limited ambiguity, and non-trivial search
spaces, a verifiable local SLM--solver pipeline should match or outperform
self-consistency while using fewer generated tokens and fewer serial model
calls. This hypothesis is intentionally conditional. It does not predict that
VFR-LLM will beat direct answering when a small model already solves the task in
one short call. The formalism is expected to matter: a bounded Datalog-like
layer with finite-domain constraints should be easier for a small local model
to produce than full first-order logic, while still being more useful than a
task-specific list of ad hoc predicates. The main bottleneck is expected to be
semantic fidelity in the natural-language-to-formalization step, rather than
solver execution.

\paragraph{Empirical status.}
The completed local matrix supports the hypothesis in several narrow strata:
Qwen pure precedence, Qwen BBH-derived pairwise and extended logical deduction,
and Gemma generated precedence. It gives mixed evidence on typed constraints:
Gemma typed improves over self-consistency but remains expensive relative to
direct answering, Qwen typed is blocked by direct-answer dominance, and Phi
typed reveals severe formalization failures. Gemma BBH-extended is marginal
against direct answering. The paper therefore treats the results as evidence
about where the method works and where the current formal target remains too
costly or too model-dependent.

\paragraph{Claim boundary.}
The study does not claim that symbolic solvers improve all LLM reasoning tasks.
The target class is limited to problems where relevant constraints can be
represented explicitly and solver search can substitute for repeated neural
generation.

\section{Theoretical Framework}

The paper rests on a simple but important separation of roles. A local SLM is
treated as a probabilistic translator from natural language into a constrained
intermediate representation. The symbolic component is treated as a
deterministic executor over that representation. The verification layer sits
between them: it asks whether the representation is grounded in the problem
statement before solver output is allowed to stand as an answer. This framing
places the work inside neuro-symbolic AI, but it avoids the stronger claim that
the neural model itself has become a reliable logical reasoner
\cite{hitzler2022neurosymbolic}.

\paragraph{Resource model.}
Let \(C_{\mathrm{dir}}\) denote the cost of a direct local-model answer,
\(C_{\mathrm{sc}}(k)\) the cost of self-consistency with \(k\) sampled answers,
and \(C_{\mathrm{vfr}}\) the cost of verifiable formalization and solver
execution. In serial execution,
\[
  C_{\mathrm{sc}}(k) \approx
  \sum_{i=1}^{k} C_{\mathrm{answer}, i} + C_{\mathrm{vote}},
\]
where \(C_{\mathrm{vote}}\) is usually small relative to generation. For the
proposed method,
\[
  C_{\mathrm{vfr}} =
  C_{\mathrm{formalize}} + C_{\mathrm{verify}} + C_{\mathrm{solve}} +
  C_{\mathrm{repair}}.
\]
On the target problem class, \(C_{\mathrm{solve}}\) and deterministic repair are
expected to be much smaller than additional neural generations, while
\(C_{\mathrm{formalize}}\) is non-negligible. The method can therefore reduce
resources only when formalization plus deterministic checking is cheaper than
the repeated generations it replaces. This is the same reason the method should
not be expected to beat direct answering on easy cases: if
\(C_{\mathrm{dir}}\) is small and accuracy is already high, the symbolic layer
adds overhead rather than saving computation.

\paragraph{Error model.}
Self-consistency assumes that individual samples are noisy and that answer
aggregation can suppress some of that noise \cite{wang2022selfconsistency}.
Adaptive self-consistency methods refine this assumption by spending fewer
samples on easier or already-stable instances
\cite{aggarwal2023adaptive,wang2025difficulty}.
VFR-LLM assumes a different failure structure. It assumes that some model
errors occur during translation into a formal representation, and that a subset
of those errors are locally detectable because they break schema constraints,
refer to unknown entities, contradict source spans, or make the solver report
unsatisfiable or unsupported constraints. The strongest case for VFR-LLM is not
that every error is repairable. It is that some frequent local-model errors are
cheaper to diagnose and repair symbolically than to resample several new
natural-language answers.

This distinction leads to a falsifiable prediction. If direct answers are
accurate and formalization errors are rare but expensive to prevent, VFR-LLM
should lose on resource metrics. If formalization errors are frequent but not
detectable from source spans or solver diagnostics, VFR-LLM should also lose,
because repair would require additional model calls or human intervention. The
method is expected to help only in the middle region: structured tasks where
the natural-language constraints are explicit, a single final answer is not
trusted, and a bounded formalization error can be caught before it becomes an
incorrect final answer.

\paragraph{Why this formalism.}
The target formalism is chosen according to four criteria: expressive adequacy,
translation reliability, bounded inference, and auditability. Full first-order
logic is expressive, but it asks a small local model to produce brittle syntax
and broad semantic commitments. A task-specific precedence encoding is easy to
solve, but it does not generalize beyond narrow ordering tasks. Datalog-style
Horn rules provide a middle ground: variables range over finite extensional
facts, rules can be checked for safety, and derivations remain inspectable
\cite{ceri1989datalog}. Finite-domain constraints then cover the assignment,
ordering, and scheduling fragments where solver search can replace repeated
model sampling. SMT and constraint solvers show why such delegation can be
effective when the formalization is faithful \cite{demoura2008z3}, but this
paper deliberately restricts the front end so that a small local model has a
realistic translation target.

\paragraph{Formal language used in this paper.}
The logical target can be described as a typed, finite-domain,
Datalog-like rule language with an attached constraint layer. A formalized
problem is a tuple
\[
  \mathcal{P} = \langle E, T, F, R, K, q \rangle,
\]
where \(E\) is a finite set of entity constants, \(T\) assigns optional types to
those constants, \(F\) is a set of grounded atoms, \(R\) is a set of safe
Horn-style rules, \(K\) is a set of finite-domain constraints, and \(q\) is the
query to be answered. The language has no function symbols and no open-ended
quantification over objects outside \(E\). Every variable in a rule must be
range-restricted by a finite type or entity predicate, so the rule can be
grounded before solving.

In the current implementation, the main predicates are deliberately small. The
predicate \(\mathrm{before}(x,y)\) means that entity \(x\) has a lower position
than entity \(y\), and \(\mathrm{after}(x,y)\) is its inverse. The predicate
\(\mathrm{position}(x,n)\) fixes entity \(x\) at the one-indexed position \(n\)
in the ordering direction requested by the question. The typed predicate
\(\mathrm{type\_before}(\tau_1,\tau_2)\) abbreviates the finite set of
constraints saying that every entity of type \(\tau_1\) precedes every entity
of type \(\tau_2\). For example, if
\(\mathrm{type}(ingest,data)\), \(\mathrm{type}(train,model)\), and
\(\mathrm{type\_before}(data,model)\) are present, the grounding step derives
\(\mathrm{before}(ingest,train)\). Solver-facing constraints then assign each
entity a finite integer position, enforce grounded precedence relations, and
enforce any fixed-position facts. This keeps ordinal statements such as
``leftmost'', ``newest'', or ``finished second'' inside the same finite-domain
semantics rather than treating them as informal natural-language hints.

This is not full first-order logic. It excludes arbitrary functions,
unbounded quantification, and broad background theories. It is also not merely
a list of ad hoc precedence edges: typed rules can compactly express group-level
constraints that expand into many item-level constraints only after grounding.
The formalism is therefore intentionally modest. Its purpose is to give local
SLMs a target that is structured enough to support verification and solver
execution, but bounded enough that translation errors can be detected and the
resulting program can be audited by a reviewer.

\paragraph{Operational semantics of the typed fragment.}
For the fragment evaluated in this paper, the semantics is defined by a finite
grounding relation followed by finite-domain model search. Let
\(\mathcal{P}=\langle E,T,F,R,K,q\rangle\). A formalization is well formed only
if every constant in \(F\), \(R\), and \(K\) belongs to the finite entity set
\(E\), every type mentioned by a typed rule occurs in the range of \(T\), and
every rule is safe, meaning that each variable is range-restricted by \(E\) or
by a finite type class. Ill-formed programs are rejected before solving.

The grounding judgment has two outputs: grounded precedence relations
\(\mathcal{P}\vdash_g \mathrm{before}(a,b)\) and fixed finite-domain positions
\(\mathcal{P}\vdash_g \mathrm{fixed}(a,n)\). It is the least relation closed
under the following rules, for \(a,b\in E\),
\(n\in\{1,\ldots,|E|\}\), and
\(\tau_1,\tau_2\in \mathrm{range}(T)\):
\[
\frac{\mathrm{before}(a,b)\in K}
     {\mathcal{P}\vdash_g \mathrm{before}(a,b)}
\qquad
\frac{\mathrm{after}(a,b)\in K}
     {\mathcal{P}\vdash_g \mathrm{before}(b,a)}
\]
\[
\frac{\mathrm{type\_before}(\tau_1,\tau_2)\in K
      \quad T(a)=\tau_1 \quad T(b)=\tau_2}
     {\mathcal{P}\vdash_g \mathrm{before}(a,b)}.
\]
\[
\frac{\mathrm{position}(a,n)\in K}
     {\mathcal{P}\vdash_g \mathrm{fixed}(a,n)}.
\]
These rules define the current deductive closure used by the implementation.
They intentionally do not materialize a separate transitive-closure predicate.
If \(\mathcal{P}\vdash_g \mathrm{before}(a,b)\) and
\(\mathcal{P}\vdash_g \mathrm{before}(b,c)\), the solver semantics implies
\(\mathrm{position}(a)<\mathrm{position}(c)\) in every satisfying assignment,
but the system does not need to add \(\mathrm{before}(a,c)\) as an explicit
derived atom.

A model of the grounded program is a bijective assignment
\[
  M:E\rightarrow \{1,\ldots,|E|\},
\]
where \(M(a)\) is the finite-domain position of entity \(a\). The assignment
satisfies the program, written \(M\models\mathcal{P}\), iff it satisfies the
all-distinct position constraint and, for every grounded relation
\(\mathcal{P}\vdash_g \mathrm{before}(a,b)\), \(M(a)<M(b)\), and for every
fixed-position judgment \(\mathcal{P}\vdash_g \mathrm{fixed}(a,n)\),
\(M(a)=n\). If no such assignment exists, the solver returns
\(\mathrm{unsat}\). If grounding fails because a rule is unsupported, a
constant is unknown, a type is empty, a fixed position is outside the finite
domain, or no constraint can be extracted, the system returns
\(\mathrm{invalid}\). Otherwise the answer to \(q\) is the total order induced
by any satisfying assignment reported by the solver.

This semantics is modest but auditable. It separates three objects: extracted
claims \((K,F,T)\), grounded relations \((\vdash_g)\), and solver-selected
models \((M\models\mathcal{P})\). A reviewer can therefore identify whether an
answer came from the local model's translation, deterministic grounding, or
finite-domain search. This separation is central to the paper's claim: the
local SLM is not credited with performing the symbolic deduction; it is
credited, when successful, with producing a representation that the symbolic
layer can check and execute.

\paragraph{Relation to tool use and executable reasoning.}
Tool-augmented and executable-reasoning methods demonstrate that language
models can gain reliability by delegating parts of a task to external
procedures
\cite{mialon2023augmented,yao2023react,schick2023toolformer,gao2023pal,chen2023program}.
The present work adopts that division of labor but adds a stricter
resource-aware criterion. Calling a tool is not counted as a success unless it
improves a measured accuracy--resource trade-off or provides an auditable trace
that direct answering cannot provide. In this sense, VFR-LLM is not simply
``LLM plus solver''. It is a budgeted translation-and-verification policy for
local SLMs.

\paragraph{Theoretical scope.}
The framework implies three regimes. In the \emph{direct-answer regime}, a small
model already answers reliably in one call; VFR-LLM should not be preferred on
resource grounds. In the \emph{repeated-sampling regime}, users would otherwise
pay for multiple local generations to stabilize an answer; VFR-LLM may reduce
calls and tokens if formalization is reliable enough. In the \emph{semantic-gap
regime}, the selected formalism cannot express the relevant meaning or the
translation is not source-faithful; the solver may make the answer look more
rigorous without making it more correct. The experiments are designed to locate
these regimes rather than to assume in advance that symbolic reasoning is
always beneficial.

\section{Related Work}

\paragraph{Self-consistency.}
Self-consistency generates multiple reasoning paths for the same problem and
selects an answer through voting or aggregation. It is useful when individual
reasoning traces are noisy, but it increases cost approximately in proportion to
the number of samples \cite{wang2022selfconsistency}.

\paragraph{Cost-aware self-consistency.}
Several variants reduce this overhead by adapting the number of sampled
reasoning paths. Adaptive-Consistency stops sampling when the answer
distribution is already stable, while Difficulty-Adaptive Self-Consistency uses
question difficulty to allocate more samples only where they are expected to be
useful \cite{aggarwal2023adaptive,wang2025difficulty}. These methods are close
to the resource question studied here, but they remain sampling policies: they
reduce repeated neural generations without replacing the reasoning step by a
verifiable symbolic computation. VFR-LLM is therefore complementary rather than
a direct substitute for adaptive sampling.

\paragraph{Resource-constrained local language models.}
Work on edge-efficient LLMs and SLMs studies how language models can be made
usable under memory, latency, energy, and hardware constraints
\cite{wang2025edge,corradini2025slmreview}. Recent local-deployment models
such as SmallThinker make these constraints explicit in the model architecture
and inference path \cite{song2025smallthinker}. This literature motivates the
hardware setting of the present study, but it usually improves efficiency
through model design, compression, inference engines, or deployment strategies.
It does not, by itself, answer whether a small local model can reduce repeated
reasoning-time generation by translating a problem into a bounded logical
representation.

\paragraph{LLM--symbolic solver systems.}
Recent neuro-symbolic systems use LLMs to translate natural-language problems
into executable representations for solvers such as SAT, SMT, Prolog, Datalog,
or theorem provers. These systems can be effective when a problem has explicit
constraints or requires backtracking. However, they remain sensitive to
translation errors, incomplete formalizations, and mismatches between the
problem statement and the solver input. This work follows the broader
neuro-symbolic tradition of combining learned perception or language
understanding with symbolic reasoning \cite{hitzler2022neurosymbolic}.

\paragraph{Executable intermediate reasoning.}
Program-aided and program-of-thought methods ask the language model to express
part of the solution as executable code, then delegate computation to an
external interpreter \cite{gao2023pal,chen2023program}. These approaches
demonstrate the value of separating language understanding from deterministic
execution. Logic-LM follows a related direction for logical reasoning by
translating natural-language problems into symbolic formulations and using
solver feedback for refinement \cite{pan2023logiclm}. The present work is
narrower in empirical scope but different in emphasis: it studies a bounded
rule-and-constraint target for small local models and makes resource accounting
against repeated sampling a first-class part of the evaluation.

\begin{table*}[htbp]
\centering
\small
\begin{tabular}{
  >{\raggedright\arraybackslash}p{0.17\textwidth}
  >{\raggedright\arraybackslash}p{0.23\textwidth}
  >{\raggedright\arraybackslash}p{0.22\textwidth}
  >{\raggedright\arraybackslash}p{0.25\textwidth}}
\toprule
Line of work & Main mechanism & Open issue for local SLM reasoning &
Position of this article \\
\midrule
Serial and adaptive self-consistency &
Generate several answer traces and aggregate, or stop sampling once the answer
distribution appears stable
\cite{wang2022selfconsistency,aggarwal2023adaptive,wang2025difficulty}. &
The reasoning step remains neural and repeated; lower sampling budgets may
reduce cost but do not make the intermediate reasoning state executable or
auditable. &
VFR-LLM replaces repeated sampled reasoning, when possible, with one
formalization call plus deterministic finite-domain solving, and evaluates
calls, tokens, latency, and accuracy against both serial and adaptive sampling. \\
\addlinespace
Program-aided and program-of-thought reasoning &
Ask the model to write executable code or program fragments and use an
interpreter for computation \cite{gao2023pal,chen2023program}. &
Execution can improve arithmetic or procedural reliability, but the generated
program may still be an unverified interpretation of the natural-language
problem. &
VFR-LLM uses a deliberately smaller typed rule-and-constraint language with
source spans, coverage checks, and deterministic repairs before execution. \\
\addlinespace
Logic-LM and LLM--solver pipelines &
Translate natural language into logical or solver-readable forms and use solver
feedback to refine the result \cite{pan2023logiclm}. &
The central evaluation question is usually logical accuracy or solver-assisted
reasoning, not whether a small local model can save repeated inference under a
measured resource budget. &
The contribution here is not the existence of an LLM--solver connection; it is
a local policy for verifiable formalization, deterministic repair, and
resource-aware comparison against repeated sampling. \\
\addlinespace
Neuro-symbolic and efficient symbolic-neural systems &
Combine learned components with symbolic representations, sometimes under data,
compute, or hardware constraints
\cite{hitzler2022neurosymbolic,li2023scallop,saha2024tinyns}. &
Many systems optimize learning pipelines, symbolic operators, or embedded
deployment, but do not isolate inference-time formalization errors made by
local language models. &
VFR-LLM focuses on the translation bottleneck at inference time: it logs
whether the model produced a faithful finite-domain formalization, whether the
solver helped, and when the symbolic layer should not be credited. \\
\bottomrule
\end{tabular}
\caption{Positioning of VFR-LLM relative to closely related approaches. The
article shares the broader neuro-symbolic premise that deterministic execution
can complement learned language understanding, but its specific contribution is
a bounded local inference policy with traceability, deterministic repair, and
explicit resource accounting against repeated sampling.}
\label{tab:prior-work-positioning}
\end{table*}

\paragraph{Efficient neurosymbolic programming.}
Neurosymbolic programming systems such as Scallop show that Datalog-style
relational representations can support interpretable and data- or
compute-efficient learning workflows \cite{li2023scallop}. Platform-aware
neurosymbolic TinyML systems such as TinyNS go further by optimizing symbolic
and neural operators under real hardware budgets \cite{saha2024tinyns}. These
systems are important precedents for resource-aware neurosymbolic design, but
they are not local SLM reasoning pipelines. The present study inherits their
hardware-conscious motivation while focusing on inference-time translation,
verification, and repair for small language models.

\paragraph{Why a bounded rule language.}
Full first-order logic is attractive as a theoretical target, but it is often a
poor engineering target for small local models: syntax is brittle, semantic
commitments are broad, and unrestricted inference can be difficult to budget.
Datalog-style Horn rules over finite domains occupy a more practical middle
ground. They support relational inference and transparent derivations while
remaining decidable, auditable, and compatible with constraint solvers for
ordering, assignment, and scheduling subproblems. This paper therefore treats
the choice of formalism as part of the contribution rather than a background
implementation detail.

\paragraph{Faithfulness and verification.}
A solver can correctly solve the wrong formal problem if the translation from
natural language is incomplete or semantically distorted. This motivates
traceable intermediate representations, round-trip checks, and diagnostic repair
mechanisms.

\section{Proposed Method}

The proposed architecture, Verifiable Formalization and Repair for LLMs
(VFR-LLM), consists of five stages. The name is intentionally literal:
\emph{verifiable} means that extracted claims must be traceable to the source
problem, \emph{formalization} is the typed finite-domain representation given
to the solver, and \emph{repair} denotes deterministic local corrections
licensed by source spans and solver diagnostics. The design is deliberately
modest: it does not assume that the language model has become a reliable
theorem prover. Instead, it treats the model as a translator into a bounded
logical language whose output must be checked before a symbolic component is
allowed to answer.

\begin{enumerate}[leftmargin=*]
  \item \textbf{Formalization.} The LLM maps the natural-language problem into a
  structured intermediate representation containing typed entities, grounded
  facts, safe Horn-style rules, finite-domain constraints, source text spans,
  and the target question.
  \item \textbf{Coverage checking.} The system verifies whether each formal
  constraint is grounded in the original text and flags missing or unsupported
  constraints.
  \item \textbf{Solver execution.} A symbolic solver executes the formal
  reasoning step and returns an answer, a model, or failure diagnostics.
  \item \textbf{Repair.} If the solver fails or diagnostics indicate incomplete
  formalization, a deterministic repair module applies local edits to the
  intermediate representation. The current implementation repairs common
  small-model errors such as non-canonical comparison syntax, inverted
  arguments that contradict a source span, and non-binary introductory rules.
  \item \textbf{Answer generation.} The final answer is produced from the solver
  result and linked back to the constraints that justify it.
\end{enumerate}

The first implementation uses Python and Pydantic for schema validation. For
the precedence tasks reported here, the symbolic backend is an exact
enumerative solver over the candidate orderings. This choice is intentional:
the current tasks are small enough that enumeration is transparent, removes a
native solver dependency from the reviewer path, and avoids attributing
implementation failures to the neuro-symbolic idea. More expressive backends
such as Datalog engines, SMT, ASP, or CP-SAT are treated as extensions of the
same finite-domain rule-and-constraint interface rather than as replacements
for the interface itself. Every result reported in this paper is produced by
this enumerative finite-domain solver. The Datalog, SMT, ASP, and CP-SAT engines
named above are not exercised in the current evaluation; they only indicate
where the same rule-and-constraint interface would connect to a standard solver.
Because the enumerative backend searches the full space of bijective position
assignments, it is a faithful executor of the finite-domain semantics for the
small entity counts evaluated here, at most seven entities, and therefore does
not understate solver difficulty on these tasks.

\paragraph{Logical target language.}
The target language is a typed finite-domain rule-and-constraint formalism. It
contains three parts:

\begin{enumerate}[leftmargin=*]
  \item \textbf{Grounded facts}, such as
  \texttt{entity(task, ingest)} or \texttt{depends\_on(index, normalize)}.
  \item \textbf{Safe Horn-style rules}, such as
  \texttt{before(X,Y) :- depends\_on(Y,X)}, where variables range over the
  finite set of extracted entities.
  \item \textbf{Solver-facing finite-domain constraints}, including pairwise
  ordering constraints, absolute one-indexed position constraints, and an
  all-distinct condition over entity positions.
\end{enumerate}

This formalism is intentionally less expressive than unrestricted first-order
logic. The restriction is a feature, not a weakness: it reduces the translation
burden for small local models, keeps inference budgetable, and gives reviewers
a compact program that can be inspected line by line.

\paragraph{Representation.}
Each formalization contains entities, typed variables, facts, rules,
constraints, source spans, a question, and metadata. The source span field is
central: it allows the system to distinguish solver failures from translation
failures and makes it possible to audit whether a formal rule is supported by
the original input. In the current implementation, each ordering instance is
compiled into a \texttt{RuleProgram} containing \texttt{entity(...)} facts, a
general rule connecting \texttt{after} and \texttt{before}, explicit
finite-domain ordering constraints, and fixed-position constraints when the
text contains ordinal or absolute-position information. The target language
therefore contains \texttt{before(A,B)}, \texttt{after(A,B)},
\texttt{type\_before(T1,T2)}, and \texttt{position(A,N)}, but not arbitrary
quantification, function symbols, or unbounded background theories.

\paragraph{Cost-aware decision policy.}
The pipeline is evaluated under an inference budget. If the first
formalization is executable and passes coverage checks, the system uses the
solver result directly. If diagnostics indicate missing or invalid constraints,
the repair module receives only the structured diagnostics rather than a full
repeat of the original reasoning task. The current repair operations are
deterministic, local, and specified before evaluation: they normalize common
comparison syntax, remove unsupported introductory rules, and repair inverted
binary arguments only when the original source span explicitly licenses the
correction. The repair module has no access to the gold answer or to aggregate
benchmark outcomes. This design is meant to make each correction inspectable;
it is not meant to hide an additional informal reasoning step behind the solver.

The resource policy is also intentionally conservative. Calls and generated
tokens are counted end to end for each method, and latency is measured for
serial execution on a single loaded local model. Parallel self-consistency could
reduce wall-clock time on machines with enough memory and throughput, so latency
claims must be interpreted under the local serial budget. Model calls and
generated tokens remain relevant because they correspond to repeated local
decoding work even when no API fee is paid.

\paragraph{Baseline separation.}
The basic SLM--solver baseline is allowed to use schema-constrained generation
and syntax normalization, so that it is not penalized for superficial forms such
as \(A > B\). VFR-LLM adds source-grounded validation and repair. This
distinction is important because the intended contribution is not merely a more
permissive parser; it is the use of traceability to detect and correct
translation errors that would otherwise be accepted by a solver.

\section{Experimental Design}

The evaluation compares five core methods:

\begin{enumerate}[leftmargin=*]
  \item Direct local SLM answer.
  \item Chain-of-thought prompting.
  \item Self-consistency with \(k=5\) sampled answers.
  \item Basic SLM--solver translation.
  \item VFR-LLM with traceable formalization, verification, and repair.
\end{enumerate}

\noindent A sixth method, cost-aware adaptive self-consistency, is evaluated on
the Qwen and Gemma strata as a stronger repeated-sampling baseline (described
below).

The benchmark focuses on ordering and finite constraint satisfaction. This is a
deliberate scope restriction: the paper tests whether a bounded symbolic layer
can replace repeated local sampling on explicit-constraint tasks, not whether
logic improves every form of language-model reasoning. The self-consistency
baseline used throughout is serial vanilla self-consistency with a fixed budget
of five samples. To test whether a cost-aware sampler erases the call and
latency advantage of VFR-LLM, we additionally evaluate an adaptive
self-consistency baseline on the Qwen and Gemma strata that stops early once the
modal answer is stable under the two-class Beta criterion of
\cite{aggarwal2023adaptive,wang2025difficulty}; it reuses the same prompt,
temperature, and five-sample cap, so only the stopping rule differs. We report
all call and latency comparisons against serial vanilla self-consistency and
report the adaptive baseline as a separate, stronger repeated-sampling
comparison.

\paragraph{Local hardware and models.}
The primary hardware condition is a MacBook Pro with Apple M3 Pro and 18 GB of
unified memory. The local backend is LM Studio's OpenAI-compatible endpoint.
The primary model is \texttt{qwen/qwen3-4b-2507}, loaded as the
Qwen3-4B-Instruct-2507 4-bit artifact. The robustness conditions use the
Phi-4-mini-reasoning and Gemma-3n-E4B 4-bit LM Studio artifacts. The main
accelerated runtime uses MLX/Metal through LM Studio. The CPU-only condition
reloads Qwen with GPU offload disabled through \texttt{lms load --gpu off}.
CPU-only is a resource baseline, not the recommended deployment configuration.

\paragraph{Benchmark strata.}
The real-model matrix contains nine local-model strata. The first is a
generated pure-precedence benchmark with 120 Qwen3-4B instances. The second is
a generated typed-precedence benchmark with 120 Qwen3-4B instances under both
MLX/Metal and CPU-only execution. The third repeats the typed-precedence
benchmark with Microsoft Phi-4-mini-reasoning as a model-family robustness
check. Gemma-3n-E4B is then evaluated on generated pure-precedence,
generated typed-precedence, and BBH-extended tasks as a third-model robustness
check. The public-data strata are derived from BIG-Bench Hard Logical Deduction
\cite{suzgun2023challenging,srivastava2023beyond}. The original pairwise
BBH-derived stratum keeps only instances whose statements can be faithfully
normalized into \texttt{before} and \texttt{after} constraints and whose
normalized constraints admit a unique total order. This produces 60 instances:
53 with three entities, six with five entities, and one with seven entities.

To avoid overstating a result on a heavily filtered public subset, the expanded
evaluation adds a second BBH-derived stratum with 120 balanced instances: 40
with three entities, 40 with five entities, and 40 with seven entities. This
extended subset preserves both pairwise relations and absolute or ordinal
position constraints, such as leftmost, newest, cheapest, or finished second,
when they can be represented as \texttt{position(A,N)}. Each BBH-derived record
stores the original source file, pinned source revision, input hash, license,
transformation rule, entity count, pairwise relation count, absolute-position
count, and the normalized constraints. These external strata are not full BBH;
they are public-source tests for the finite-domain fragment claimed by the
paper. The deterministic formalism ablations are larger: 600 pure-precedence
instances, 300 typed-precedence instances, and the 120-instance extended BBH
formalism ablation. These ablations separate solver expressivity from model
translation quality.

\paragraph{Prompting protocol.}
All local SLM calls use LM Studio's OpenAI-compatible chat endpoint with
schema-constrained JSON output. Table \ref{tab:prompting-protocol} summarizes
the prompts used by each method; the appendix gives the executable templates
and decoding parameters. Direct answering,
chain-of-thought, and self-consistency all return the same two-field answer
schema, \texttt{final\_answer} and \texttt{reasoning}. Basic SLM--solver and
VFR-LLM use the same formalization prompt and the same JSON schema; the
difference is post-generation handling. The basic solver receives only syntax
normalization, whereas VFR-LLM adds source-grounded validation, deterministic
repair, and coverage diagnostics before solver execution. No method receives
the gold answer, aggregate benchmark statistics, or information about which
baseline is being evaluated.

\paragraph{Randomness and decoding control.}
The deterministic parts of the artifact use explicit seeds: 13 for the
generated pure-precedence benchmark, 29 for the generated typed-precedence
benchmark, 101 for the pairwise BBH-derived subset, 211 for the extended
BBH-derived subset, 503 for the disjoint heldout subset, and 13 for the audit
sample and bootstrap resampling offsets. Local SLM calls are controlled by
fixed model identifiers, prompts, JSON schemas, token budgets, temperatures,
and output paths. Direct answering, chain-of-thought, formalization, and
VFR-LLM use temperature 0.0; self-consistency and adaptive self-consistency use
temperature 0.7. The LM Studio backend used for the reported runs does not
expose or enforce a backend-level random seed in this artifact. The local-model
results should therefore be read as protocol-reproducible rather than
bit-for-bit reproducible: a rerun under the same protocol may vary slightly
because of backend decoding and runtime nondeterminism.

\begin{table*}[htbp]
\centering
\small
\begin{tabular}{
  >{\raggedright\arraybackslash}p{0.17\textwidth}
  >{\raggedright\arraybackslash}p{0.24\textwidth}
  >{\raggedright\arraybackslash}p{0.18\textwidth}
  >{\raggedright\arraybackslash}p{0.29\textwidth}}
\toprule
Method & User instruction & Decoding & Parsed output and post-processing \\
\midrule
Direct answer &
Solve the problem and give the order from first to last. &
\(T=0.0\), 128 answer tokens, one call. &
JSON answer schema. The predicted order is normalized and compared with the
expected order. \\
Chain-of-thought &
Solve step by step, then give the final order from first to last. &
\(T=0.0\), 128 answer tokens, one call. &
Same answer schema as direct answering; only the prompting instruction differs. \\
Self-consistency &
Same step-by-step prompt as chain-of-thought. &
\(k=5\), \(T=0.7\), 128 answer tokens per sample. &
Five parsed answers are normalized and aggregated by majority vote. \\
Basic SLM--solver &
Formalize the problem with source spans for every constraint. &
\(T=0.0\), 1024--1280 formalization tokens, one call. &
JSON formalization schema. Syntax is normalized, then the solver is run without
source-grounded repair. \\
VFR-LLM &
Same formalization prompt as the basic SLM--solver. &
\(T=0.0\), 1024--1280 formalization tokens, one call. &
JSON formalization schema, followed by source-grounded validation,
deterministic repair, coverage diagnostics, and solver execution. \\
\bottomrule
\end{tabular}
\caption{Prompting protocol for the local-model experiments. The formalization
token cap is 1024 by default and 1280 for the extended or longer
formalization conditions specified in the experiment matrix.}
\label{tab:prompting-protocol}
\end{table*}

\paragraph{Decision rule.}
The paper treats a resource-reduction claim as supported only if VFR-LLM is
competitive with serial self-consistency in accuracy, reduces calls or
generated tokens, and is not dominated by direct answering in the same stratum.
The interpretation is stricter when direct answering is much cheaper and VFR's
accuracy gain over direct answering is below five percentage points; such cases
are reported as marginal even when VFR saves calls relative to
self-consistency. This decision rule is intentionally conservative. If VFR-LLM
improves auditability or failure diagnosis but is less accurate or more
expensive than direct answering, the result is reported as a reliability or
diagnostic result, not as a compute-reduction result.

\paragraph{Metrics.}
The main metrics are end-to-end accuracy, number of SLM calls, prompt tokens,
completion tokens, total tokens, serial wall-clock latency, model-only latency,
solver latency, solver executability, solver status, repair attempts, process
RSS proxies, and failure mode. Accuracy is reported with Wilson 95\%
intervals. Token and latency means are reported with bootstrap intervals over
problem instances, and paired comparisons against direct answering,
chain-of-thought, self-consistency, and the basic SLM--solver baseline use
problem-level matched differences. Accuracy disagreements are summarized with
exact McNemar tests. The current artifact does not measure energy and does not
provide low-level memory profiling, so no energy claim is made.

\paragraph{Cost interpretation.}
Because the target deployment is a local machine, monetary API cost is not the
primary resource. The reported proxies are model invocations, generated tokens,
serial wall-clock latency, and logged process-level RSS metadata. Mean values
and per-problem standard deviations are reported for the BBH-extended resource
comparison because that condition carries the strongest latency claim. A method
is described as resource-saving only with respect to the measured proxy on
which it improves. Fewer model calls, for example, do not by themselves imply
lower latency, lower memory pressure, or lower energy use.

\paragraph{Failure analysis.}
Every unsuccessful run is assigned to an auditable category: schema or syntax
error, unknown entity, unsupported rule, unsatisfiable constraints, semantic
mismatch, missing answer, or unclassified failure. This taxonomy avoids
treating all failures as generic model mistakes. It also distinguishes errors
that can be repaired locally from errors that require a better prompt, a
smaller formal target, a different model, or a broader benchmark redesign.

\paragraph{Automatic traceability checks.}
To reduce dependence on subjective inspection, the evaluation computes
source-grounded formalization checks from logged model outputs. For each
formalization-bearing method, the checker measures entity precision and recall
against the benchmark entity set, exact or normalized source-span coverage for
each extracted constraint, support of each formal rule by the finite-domain
language, and precision/recall of normalized constraints against the
deterministic benchmark formalizer. These metrics are not treated as a
substitute for human semantic annotation. They are used as automatic
falsification checks: a claimed improvement is less credible when the generated
formalization is ungrounded, unsupported, or missing benchmark constraints.

\paragraph{Negative controls.}
The deterministic solver is also evaluated on four deliberately invalid or
unsatisfiable controls: a cyclic precedence pair, an out-of-domain absolute
position, conflicting fixed positions, and a typed rule whose left type has no
grounded entities. These controls do not test LLM translation quality. They
test whether the symbolic layer rejects impossible formal states instead of
returning a plausible order.

\paragraph{Additional automatic robustness checks.}
After the main matrix, three non-manual robustness analyses are run. First, a
pipeline ablation compares the basic solver, a traceability-only filter that
abstains on non-traceable formalizations, and the full VFR pipeline. Second, a
disjoint BBH-derived heldout set is generated by excluding source hashes from
the main 120-instance extended subset and rerunning the deterministic
formalism ablation. Third, a repeated local-model probe reruns direct
answering, serial self-consistency, and VFR-LLM three times on the first twelve
BBH-extended instances for Qwen and Gemma. This repeated probe is used only as
a stability and latency sanity check under repeated local calls; it is not
treated as a replacement for the main 120-instance matrix.

\paragraph{Formalism selection criterion.}
A candidate formalism is retained only if it improves at least one of three
quantities without unacceptable accuracy loss: fewer model calls than
self-consistency, fewer generated tokens than repeated sampling, or higher
formalization executability than a less structured baseline. A more expressive
logic is not automatically better if the local model cannot translate into it
reliably.

\section{Results}

The completed evaluation includes deterministic symbolic ablations and nine
local-model conditions. The empirical picture is not a uniform win for the
symbolic layer. It is a set of boundary conditions. VFR-LLM is consistently
useful when direct local answering is weak and the constraints can be expressed
in the finite-domain target language. It is not a general way to make every
local query cheaper than a direct answer.

The strongest new result is the expanded BBH-derived condition with Qwen. This
condition is harder than the earlier 60-item pairwise subset: it contains 120
public-source instances balanced across three, five, and seven entities, and it
retains ordinal or absolute-position statements as \texttt{position(A,N)}
constraints. VFR-LLM reaches 0.933 accuracy, compared with 0.317 for direct
answering and 0.283 for serial self-consistency. The paired comparison against
self-consistency has 80 VFR-only correct cases and two self-consistency-only
correct cases. The exact McNemar \(p\)-value is \(1.41\times 10^{-21}\). VFR
uses one model call instead of five and is faster in serial wall-clock time,
but it does not reduce total tokens in this condition: the formalization is
longer than the short sampled answers. This matters for interpretation. The
resource benefit here is reduced repeated sampling and latency, not token
reduction.

\begin{table*}[htbp]
\centering
\resizebox{\textwidth}{!}{
\begin{tabular}{lrrrrrr}
\toprule
Method & Accuracy & Calls & Total tokens & Latency (s) & Repairs & Executable \\
\midrule
Direct LLM & 0.317 & 1.0 & 258.8 & 3.57 & 0.0 & n/a \\
Chain-of-thought & 0.283 & 1.0 & 272.1 & 3.43 & 0.0 & n/a \\
Self-consistency \(k=5\) & 0.283 & 5.0 & 618.6 & 16.54 & 0.0 & n/a \\
Basic SLM--solver & 0.275 & 1.0 & 702.9 & 11.57 & 0.0 & 0.608 \\
VFR-LLM & 0.933 & 1.0 & 702.9 & 11.32 & 3.2 & 0.983 \\
\bottomrule
\end{tabular}
}
\caption{Qwen3-4B on the 120-instance BBH-derived extended logical-deduction
subset. VFR-LLM strongly improves accuracy and reduces calls and serial
latency relative to self-consistency, but it does not reduce total tokens.}
\label{tab:qwen-bbh-extended}
\end{table*}

The difficulty split strengthens the Qwen result. Direct answering and
self-consistency degrade sharply on seven-entity instances, while VFR-LLM
remains above 0.90 accuracy in every entity-count stratum. The larger cells are
still only 40 examples each, so this is not a full scaling law. It is enough,
however, to answer the most direct objection to the earlier pairwise subset:
the positive Qwen result is not confined to three-entity cases.

\begin{table*}[htbp]
\centering
\resizebox{\textwidth}{!}{
\begin{tabular}{llrrrr}
\toprule
Model/benchmark & Entities & Direct & Self-consistency & VFR-LLM & VFR executable \\
\midrule
Qwen BBH-extended & 3 & 0.650 & 0.600 & 0.950 & 0.975 \\
Qwen BBH-extended & 5 & 0.275 & 0.250 & 0.925 & 1.000 \\
Qwen BBH-extended & 7 & 0.025 & 0.000 & 0.925 & 0.975 \\
Gemma BBH-extended & 3 & 0.700 & 0.700 & 0.600 & 1.000 \\
Gemma BBH-extended & 5 & 0.350 & 0.325 & 0.350 & 1.000 \\
Gemma BBH-extended & 7 & 0.000 & 0.025 & 0.175 & 1.000 \\
\bottomrule
\end{tabular}
}
\caption{Descriptive entity-count split on the extended public-source
benchmark. Qwen shows a strong VFR advantage across all sizes. Gemma shows only
a modest advantage on the hardest seven-entity cases and no advantage on the
easiest cases.}
\label{tab:extended-difficulty}
\end{table*}

Gemma provides a useful robustness check. It does not reproduce the magnitude
of the Qwen BBH-extended result. On generated pure-precedence tasks, VFR-LLM is
substantially more accurate than direct answering and self-consistency while
using one call and fewer tokens than self-consistency. On generated typed tasks,
VFR-LLM improves over direct answering and self-consistency, but only slightly
over the basic solver, and it remains much slower than direct answering. On
the BBH-extended public-source condition, VFR-LLM reaches 0.375 accuracy versus
0.350 for direct answering and self-consistency. The paired comparison against
direct answering is not significant (\(p=0.72\)). This condition should
therefore be treated as marginal support for reduced repeated sampling, not as
strong evidence that Gemma benefits from the full VFR pipeline.

\begin{table*}[htbp]
\centering
\resizebox{\textwidth}{!}{
\begin{tabular}{llrrrrrl}
\toprule
Condition & Model & Direct & SC & Basic & VFR & Token reduction vs. SC & Decision \\
\midrule
Precedence & Qwen & 0.675 & 0.700 & 0.483 & 0.983 & 34.4\% & Strong support \\
Typed & Qwen & 0.850 & 0.833 & 0.750 & 0.817 & 39.7\% & Blocked by direct \\
BBH pairwise & Qwen & 0.850 & 0.867 & 0.317 & 0.967 & 49.2\% & Strong support \\
BBH extended & Qwen & 0.317 & 0.283 & 0.275 & 0.933 & -13.6\% & Calls/latency support \\
Typed CPU & Qwen & 0.850 & 0.833 & 0.750 & 0.817 & 40.2\% & Blocked by direct \\
Typed & Phi & 0.183 & 0.133 & 0.008 & 0.042 & 40.9\% & Negative \\
Precedence & Gemma & 0.375 & 0.367 & 0.375 & 0.683 & 39.1\% & Support vs. SC \\
Typed & Gemma & 0.408 & 0.433 & 0.575 & 0.608 & 31.5\% & Mixed support \\
BBH extended & Gemma & 0.350 & 0.350 & 0.108 & 0.375 & 49.5\% & Marginal vs. direct \\
\bottomrule
\end{tabular}
}
\caption{Decision-oriented summary of the local-model matrix. ``SC'' denotes
self-consistency with \(k=5\). The decision column is interpretive: it combines
accuracy, resource proxies, direct-answer dominance, and paired comparisons.}
\label{tab:decision-summary}
\end{table*}

Table~\ref{tab:claim-support} makes the scope of the empirical claims explicit.
This is important because several superficially attractive interpretations are
not supported by the evidence. The results support replacing repeated sampling
in bounded ordering-like tasks when the model can produce an executable
formalization. They do not support a model-agnostic or hardware-level claim that
adding a symbolic layer always makes local inference cheaper.

\begin{table*}[htbp]
\centering
\resizebox{\textwidth}{!}{
\begin{tabular}{>{\raggedright\arraybackslash}p{0.22\textwidth}>{\raggedright\arraybackslash}p{0.24\textwidth}>{\raggedright\arraybackslash}p{0.24\textwidth}>{\raggedright\arraybackslash}p{0.24\textwidth}}
\toprule
Claim & Supporting evidence & Limiting evidence & Interpretation used here \\
\midrule
VFR-LLM can replace serial self-consistency on bounded ordering tasks &
Qwen precedence, Qwen BBH pairwise, Qwen BBH extended, and Gemma precedence
all improve over self-consistency while using one model call. &
Qwen typed is dominated by direct answering; Phi typed is negative; Gemma
BBH extended is only marginal against direct answering. &
Supported only for finite-domain structured tasks where direct answering is
weak and formalization is faithful. \\
The extended finite-domain formalism is needed for the public-source subset &
The deterministic BBH-extended ablation gives 1.000 accuracy for the typed
finite-domain solver and 0.208 for the precedence-only solver. &
Solver capacity is not the main bottleneck in local-model runs; translation
errors still determine whether the formalism helps. &
The formalism is justified as an executable target, not as a guarantee that a
small model will always translate into it correctly. \\
VFR-LLM reduces local computation &
It reduces model calls relative to \(k=5\) self-consistency and reduces serial
latency on Qwen and Gemma BBH-extended; most generated-task conditions also
reduce tokens relative to self-consistency. &
Qwen BBH-extended uses more total tokens than self-consistency, direct
answering remains cheaper in several typed conditions, and energy is not
measured. &
The claim is proxy-specific: reduced repeated sampling and, in selected
strata, reduced latency or tokens. \\
VFR-LLM is a general-purpose compute reducer for all local SLM use &
No condition supports this broad statement. &
The matrix contains negative, blocked, and marginal cases across model
families and benchmarks. &
Not claimed. The contribution is a bounded accuracy--resource trade-off for
structured finite-domain reasoning. \\
\bottomrule
\end{tabular}
}
\caption{Empirical scope of the main claims. The table separates supported
claims from interpretations that the completed matrix does not justify.}
\label{tab:claim-support}
\end{table*}

The deterministic ablations explain why the extended formalism is not merely a
parser convenience. On the 120-instance BBH-extended deterministic ablation,
the typed finite-domain solver solves every instance. A precedence-only solver,
which ignores \texttt{position(A,N)}, solves only 0.208 of the same instances.
The text-order heuristic reaches 0.067. Thus fixed-position constraints are
needed for this public-source subset. The bottleneck in the real-model runs is
not solver expressivity. It is whether a local model can produce or support a
faithful formalization cheaply enough.

\begin{table*}[htbp]
\centering
\begin{tabular}{lrrr}
\toprule
Deterministic method & \(n\) & Accuracy & Executable rate \\
\midrule
Text-order heuristic & 120 & 0.067 & n/a \\
Precedence-only solver & 120 & 0.208 & 0.208 \\
Typed finite-domain solver & 120 & 1.000 & 1.000 \\
\bottomrule
\end{tabular}
\caption{Formalism ablation on the BBH-derived extended benchmark. Absolute
or ordinal position constraints are necessary for this subset.}
\label{tab:extended-formalism-ablation}
\end{table*}

The automatic traceability checks help explain the model-family difference.
On Qwen BBH-extended, VFR-LLM preserves nearly complete entity coverage,
source-span grounding, supported-rule syntax, and normalized constraint recall.
The fully traceable rate is 0.942, compared with 0.258 for the basic
SLM--solver. On Gemma, the checker is more cautionary: VFR-LLM has high
source-span and rule-support rates, but normalized constraint recall is only
0.496 and the fully traceable rate is 0.292. This automatic result is
consistent with the weaker Gemma accuracy result and supports the paper's
bounded interpretation: the symbolic layer helps only when the local model can
translate the relevant constraints faithfully.

\begin{table*}[htbp]
\centering
\resizebox{\textwidth}{!}{
\begin{tabular}{llrrrrr}
\toprule
Model & Method & Entity recall & Source coverage & Rule support & Constraint recall & Fully traceable \\
\midrule
Qwen & Basic SLM--solver & 0.991 & 0.988 & 0.976 & 0.475 & 0.258 \\
Qwen & VFR-LLM & 0.991 & 0.998 & 0.990 & 0.982 & 0.942 \\
Gemma & Basic SLM--solver & 0.985 & 0.983 & 0.983 & 0.293 & 0.058 \\
Gemma & VFR-LLM & 0.985 & 0.988 & 0.988 & 0.496 & 0.292 \\
\bottomrule
\end{tabular}
}
\caption{Automatic formalization-quality checks on the BBH-extended runs.
Constraint recall compares normalized generated rules against the deterministic
benchmark formalizer. ``Fully traceable'' requires full entity recall,
perfect entity and constraint precision, source-grounded constraints, supported
formal rules, and complete normalized constraint recall for a problem.}
\label{tab:formalization-quality}
\end{table*}

The negative controls provide a separate sanity check on the symbolic layer.
Both deterministic solver variants reject every deliberately impossible
instance. The typed solver classifies two cases as unsatisfiable and two as
invalid, while the precedence-only solver rejects one as unsatisfiable and
three as invalid because it intentionally lacks \texttt{position} and
\texttt{type\_before} support. These controls do not establish that LLM
formalizations are semantically correct, but they reduce a different risk: the
solver is not silently accepting impossible formal states.

\begin{table}[htbp]
\centering
\small
\begin{tabular}{lrrrr}
\toprule
Method & \(n\) & Rejected & Unsat & Invalid \\
\midrule
Prec.-only solver & 4 & 1.000 & 1 & 3 \\
Typed FD solver & 4 & 1.000 & 2 & 2 \\
\bottomrule
\end{tabular}
\caption{Deterministic negative controls. Rejected means the solver returned
\texttt{unsat} or \texttt{invalid} rather than a plausible order.}
\label{tab:negative-controls}
\end{table}

The pipeline ablation clarifies what VFR adds beyond simply passing an LLM
formalization to a solver. On Qwen BBH-extended, the basic solver reaches only
0.275 accuracy. A traceability-only filter is conservative: it accepts 0.258 of
formalizations and reaches 0.250 end-to-end accuracy if abstentions are counted
as incorrect. The full VFR pipeline reaches 0.933 accuracy and 0.983 solver
executability. Gemma shows the same ordering but a weaker absolute result:
full VFR improves over the basic solver, yet only reaches 0.375 accuracy. The
ablation therefore supports the necessity of both traceability and repair for
the Qwen result, while reinforcing the model-family boundary.

\begin{table*}[htbp]
\centering
\resizebox{\textwidth}{!}{
\begin{tabular}{llrrrr}
\toprule
Model & Variant & Accuracy & Executable & Traceability accept & Abstention \\
\midrule
Qwen & Basic solver & 0.275 & 0.608 & n/a & 0.000 \\
Qwen & Traceability filter, no repair & 0.250 & 0.608 & 0.258 & 0.742 \\
Qwen & Full VFR & 0.933 & 0.983 & 0.942 & 0.000 \\
Gemma & Basic solver & 0.108 & 0.933 & n/a & 0.000 \\
Gemma & Traceability filter, no repair & 0.058 & 0.933 & 0.058 & 0.942 \\
Gemma & Full VFR & 0.375 & 1.000 & 0.292 & 0.000 \\
\bottomrule
\end{tabular}
}
\caption{Post-hoc VFR pipeline ablation on BBH-extended runs. The
traceability-only variant abstains on non-fully-traceable formalizations and
counts abstentions as incorrect for end-to-end accuracy.}
\label{tab:pipeline-ablation}
\end{table*}

Three additional automatic checks were added to reduce dependence on a single
positive table. A disjoint 60-instance BBH-derived heldout set preserves the
formalism result: the typed finite-domain solver reaches 1.000 accuracy,
whereas the precedence-only solver reaches 0.167 and the text-order heuristic
0.050. Traceability is also strongly associated with correctness. For VFR-LLM,
fully traceable Qwen formalizations are correct in 0.991 of cases, compared
with 0.000 when not fully traceable; the corresponding Gemma rates are 1.000
and 0.118. Finally, the repeated local probe supports the Qwen latency and
accuracy pattern but cautions against overgeneralizing Gemma. Across three
twelve-problem repeats, Qwen VFR-LLM remains 1.000 accurate and faster than
serial self-consistency, although it uses more tokens. Gemma VFR-LLM remains
only 0.167 accurate and is slower than both Gemma baselines on the same probe.

\begin{table*}[htbp]
\centering
\resizebox{\textwidth}{!}{
\begin{tabular}{llrrrr}
\toprule
Check & Condition & \(n\) & Key baseline & VFR or typed result & Interpretation \\
\midrule
Heldout formalism & BBH-heldout typed solver & 60 & Precedence-only 0.167 & Typed solver 1.000 & Formalism generalizes to disjoint source hashes \\
Traceability link & Qwen VFR & 120 & Not traceable 0.000 & Traceable 0.991 & Traceability predicts correctness \\
Traceability link & Gemma VFR & 120 & Not traceable 0.118 & Traceable 1.000 & Same pattern, lower traceable rate \\
Repeated probe & Qwen, 3 repeats \(\times\) 12 cases & 36 & SC \(0.278\pm0.048\), 20.92 s & VFR \(1.000\pm0.000\), 13.55 s & Supports Qwen stability; tokens are not reduced \\
Repeated probe & Gemma, 3 repeats \(\times\) 12 cases & 36 & SC \(0.361\pm0.048\), 14.84 s & VFR \(0.167\pm0.000\), 45.52 s & Confirms a negative Gemma probe \\
\bottomrule
\end{tabular}
}
\caption{Additional automatic robustness checks. The repeated probe reports
mean accuracy and mean latency across three independent local reruns; it is
not a replacement for the main 120-instance matrix.}
\label{tab:automatic-robustness}
\end{table*}

The resource summaries give the same mixed answer. Solver time is negligible
relative to model time: the measured solver component is on the order of
milliseconds per problem. The cost is the formalization call. On Qwen
BBH-extended, VFR-LLM reduces mean serial latency from 16.54 seconds for
self-consistency to 11.32 seconds, but total tokens increase from 618.6 to
702.9. On Gemma BBH-extended, VFR-LLM reduces mean latency from 17.40 seconds
to 10.88 seconds and total tokens from 1155.4 to 583.5, but the accuracy gain
over direct answering is only 0.025. The evidence therefore does not support a
uniform compute-reduction claim. The defensible interpretation is narrower: the
method can reduce repeated local sampling when the formalization is faithful,
but the benefit depends on model family and task stratum.

\begin{table*}[htbp]
\centering
\resizebox{\textwidth}{!}{
\begin{tabular}{llrrrr}
\toprule
Model & Method & Model latency (s) & LM Studio RSS (MB) & Runner RSS (MB) & Solver latency (ms) \\
\midrule
Qwen & Direct LLM & \(3.57\pm3.32\) & \(1556.7\pm477.9\) & \(84.9\pm0.0\) & n/a \\
Qwen & Self-consistency \(k=5\) & \(16.54\pm7.04\) & \(1576.2\pm426.9\) & \(84.9\pm0.0\) & n/a \\
Qwen & VFR-LLM & \(11.32\pm11.69\) & \(1533.4\pm506.9\) & \(84.9\pm0.0\) & \(2.93\pm5.44\) \\
Gemma & Direct LLM & \(3.57\pm0.72\) & \(1542.9\pm508.6\) & \(85.1\pm0.0\) & n/a \\
Gemma & Self-consistency \(k=5\) & \(17.40\pm3.63\) & \(1356.1\pm323.4\) & \(85.1\pm0.0\) & n/a \\
Gemma & VFR-LLM & \(10.88\pm2.93\) & \(1598.1\pm567.9\) & \(85.1\pm0.0\) & \(1.06\pm2.30\) \\
\bottomrule
\end{tabular}
}
\caption{Per-problem resource variability on the BBH-extended runs. Entries are
mean \(\pm\) sample standard deviation computed from logged runtime metadata.
The solver component is reported in milliseconds; RSS values are process-level
proxies rather than controlled peak-memory measurements.}
\label{tab:resource-variability}
\end{table*}

The variability table sharpens the resource interpretation. The latency
comparison against serial self-consistency is visible for both Qwen and Gemma,
but memory should not be read as a stable advantage for VFR-LLM. LM Studio RSS
varies substantially across rows, and the Python runner RSS is effectively
constant because the runner only orchestrates calls and solver checks. These
measurements support a call and latency claim against repeated sampling; they
do not support a strong memory or energy claim.

\paragraph{Adaptive self-consistency baseline.}
A natural objection is that the serial \(k=5\) self-consistency baseline is
needlessly expensive, and that a cost-aware sampler would erase the call and
latency advantage of VFR-LLM. To answer this directly, we add an adaptive
self-consistency baseline on the four Qwen strata and the three Gemma strata. It
reuses the same
chain-of-thought prompt and sampling temperature as vanilla self-consistency and
caps the budget at the same five samples, but it stops early once the modal
answer is stable under the two-class Beta criterion of Adaptive-Consistency
\cite{aggarwal2023adaptive}, evaluated exactly with posterior probability
\(0.95\). This is the strongest repeated-sampling baseline considered here.
Table~\ref{tab:adaptive-self-consistency} reports the result. Adaptive sampling
behaves as expected: it reduces the mean number of calls from \(5.0\) to between
\(4.08\) and \(4.57\) and reduces mean tokens by roughly \(15\)--\(17\%\) while
tracking the accuracy of vanilla self-consistency within \(0.016\). It does not
close the gap with VFR-LLM. On pure precedence, VFR-LLM remains far more accurate
(\(0.983\) versus \(0.700\)) using one model call instead of \(4.22\); the paired
McNemar test has 36 VFR-only and two adaptive-only correct cases
(\(p=5.4\times10^{-9}\)). On the BBH-extended subset the gap is larger still
(\(0.933\) versus \(0.283\), one call versus \(4.57\); 80 VFR-only and two
adaptive-only correct cases, \(p=1.4\times10^{-21}\)). The two strata where the
paper already declines to claim an advantage are unchanged: on BBH pairwise the
VFR--adaptive difference is not significant (\(p=0.18\)), and on typed tasks it
is not significant (\(p=1.0\)), consistent with direct-answer dominance there.
The cost-aware baseline therefore strengthens rather than weakens the bounded
claim: even against an adaptive sampler that issues roughly four calls, VFR-LLM
delivers its accuracy gains on the precedence and BBH-extended strata with a
single model call.

The Gemma replication tells the same model-dependent story and is reported in
the lower block of Table~\ref{tab:adaptive-self-consistency}. Adaptive sampling
again lowers calls from \(5.0\) to between \(4.22\) and \(4.35\) at accuracy
comparable to vanilla self-consistency. The paired VFR comparison reproduces the
boundaries the paper already draws for Gemma rather than overturning them: VFR-LLM
is strongly better on pure precedence (\(0.683\) versus \(0.367\), one call
versus \(4.30\); 46 VFR-only and eight adaptive-only correct cases,
\(p=1.4\times10^{-7}\)), only marginally better on typed tasks (\(0.608\) versus
\(0.467\), \(p=0.046\)), and statistically indistinguishable on the BBH-extended
subset (\(0.375\) versus \(0.350\), \(p=0.73\)). The adaptive baseline thus does
not rescue Gemma on the stratum where the paper already records only marginal
support, and it does not erase the clear Gemma precedence advantage. Across both
model families, the stronger repeated-sampling baseline confirms the bounded
claim instead of dissolving it.

\begin{table*}[htbp]
\centering
\resizebox{\textwidth}{!}{
\begin{tabular}{lllrrrrr}
\toprule
Model & Stratum & Method & Accuracy & Calls & Total tokens & Latency (s) & McNemar vs. VFR \\
\midrule
Qwen & Precedence & Self-consistency \(k=5\) & 0.700 & 5.00 & 839.0 & 11.57 & --- \\
Qwen & Precedence & Adaptive self-consistency & 0.700 & 4.22 & 698.6 & 9.36 & \(5.4\times10^{-9}\) \\
Qwen & Precedence & VFR-LLM & 0.983 & 1.00 & 550.0 & 8.27 & --- \\
Qwen & BBH pairwise & Self-consistency \(k=5\) & 0.867 & 5.00 & 973.0 & 9.41 & --- \\
Qwen & BBH pairwise & Adaptive self-consistency & 0.883 & 4.08 & 781.7 & 7.58 & 0.18 \\
Qwen & BBH pairwise & VFR-LLM & 0.967 & 1.00 & 494.3 & 6.55 & --- \\
Qwen & BBH extended & Self-consistency \(k=5\) & 0.283 & 5.00 & 618.6 & 16.54 & --- \\
Qwen & BBH extended & Adaptive self-consistency & 0.283 & 4.57 & 526.4 & 15.16 & \(1.4\times10^{-21}\) \\
Qwen & BBH extended & VFR-LLM & 0.933 & 1.00 & 702.9 & 11.32 & --- \\
Qwen & Typed & Self-consistency \(k=5\) & 0.833 & 5.00 & 1411.0 & 14.14 & --- \\
Qwen & Typed & Adaptive self-consistency & 0.825 & 4.08 & 1171.5 & 10.13 & 1.00 \\
Qwen & Typed & VFR-LLM & 0.817 & 1.00 & 850.4 & 16.07 & --- \\
\midrule
Gemma & Precedence & Self-consistency \(k=5\) & 0.367 & 5.00 & 928.6 & 14.05 & --- \\
Gemma & Precedence & Adaptive self-consistency & 0.367 & 4.30 & 826.0 & 7.02 & \(1.4\times10^{-7}\) \\
Gemma & Precedence & VFR-LLM & 0.683 & 1.00 & 565.9 & 10.89 & --- \\
Gemma & BBH extended & Self-consistency \(k=5\) & 0.350 & 5.00 & 1155.4 & 17.40 & --- \\
Gemma & BBH extended & Adaptive self-consistency & 0.350 & 4.35 & 1041.8 & 8.43 & 0.73 \\
Gemma & BBH extended & VFR-LLM & 0.375 & 1.00 & 583.5 & 10.88 & --- \\
Gemma & Typed & Self-consistency \(k=5\) & 0.433 & 5.00 & 1374.1 & 17.94 & --- \\
Gemma & Typed & Adaptive self-consistency & 0.467 & 4.22 & 1186.7 & 8.69 & 0.046 \\
Gemma & Typed & VFR-LLM & 0.608 & 1.00 & 941.1 & 23.36 & --- \\
\bottomrule
\end{tabular}
}
\caption{Cost-aware adaptive self-consistency on the Qwen and Gemma strata, using
the two-class Beta stopping rule of \cite{aggarwal2023adaptive} at posterior
probability \(0.95\) with a five-sample cap. Adaptive sampling lowers calls and
tokens relative to vanilla \(k=5\) self-consistency at comparable accuracy, but
the single-call accuracy advantage of VFR-LLM persists on the precedence and
Qwen BBH-extended strata. The final column gives the exact McNemar \(p\)-value of
the paired VFR-LLM versus adaptive comparison; the non-significant cells (Qwen
BBH-pairwise and typed, Gemma BBH-extended) match strata where the paper already
declines to claim an advantage.}
\label{tab:adaptive-self-consistency}
\end{table*}

Taken together, the results support a bounded and model-dependent conclusion.
Qwen provides the clearest positive evidence: VFR-LLM improves pure precedence
tasks and both BBH-derived public-source subsets while reducing repeated
sampling relative to serial self-consistency. Gemma supports the precedence
result, but its gains on typed constraints are weaker and its extended
public-source results are mixed. Phi does not support a positive typed-constraint
claim under the tested configuration. VFR-LLM is therefore best understood as a
resource-aware and auditable alternative to serial self-consistency for bounded
structured reasoning, rather than as a general-purpose compute reducer for all
local SLM use.

\section{Limitations}

The proposed approach is not expected to help equally across all reasoning
tasks. It is best suited for problems whose constraints can be explicitly
represented. Tasks that require extensive implicit knowledge, ambiguous
interpretation, or subjective judgment may require probabilistic logic,
argumentation frameworks, or human clarification.

The completed local evaluation matrix goes beyond a pilot study, but remains
bounded. It uses generated ordering-style benchmarks, two BBH-derived
public-data subsets, three local model families, and one consumer laptop. The
extended BBH-derived subset reduces the risk that the positive result is purely
synthetic or confined to three-entity pairwise problems, but it is still not the
full BBH Logical Deduction task. It keeps only cases that can be represented in
the finite-domain ordering language used here. The results therefore support a
narrow claim about local structured reasoning under this benchmark design. They
do not establish that the same trade-off will hold for arbitrary planning,
legal reasoning, or open-domain question answering. A direct next test of
generalization is to apply the same typed finite-domain interface to a
non-ordering family, such as resource assignment or single-machine scheduling,
where the solver searches over assignments rather than total orders. That
extension reuses the formalization, verification, and repair stages unchanged,
but requires a new generated or public-source benchmark and additional
local-model runs, so it is left to future work rather than reported here.

The central empirical limitation is also the central scientific result. On
pure-precedence tasks, VFR-LLM improves accuracy and reduces calls and tokens
relative to serial self-consistency for Qwen and Gemma. On Qwen typed tasks,
direct answering is more accurate and cheaper than VFR-LLM; on Gemma typed
tasks VFR improves accuracy but remains substantially slower than direct
answering; and Phi fails to translate reliably into the typed formalism. On the
BBH-derived extended subset, Qwen strongly supports VFR, but Gemma provides
only marginal evidence against direct answering. The current typed layer is
therefore not a general compute-saving mechanism. It is better interpreted as a
diagnostic instrument that reveals where formalization overhead, label
fidelity, and model capability erase the expected savings.

The method also depends on the quality of the formal vocabulary. If the
selected rule-and-constraint language cannot express the relevant semantics,
the verification layer may produce a false sense of rigor. If the language is
too expressive, the local model may fail to translate into it reliably and the
solver may become harder to budget. The deterministic ablations confirm this
trade-off: pure precedence does not require the typed rule layer, while typed
group-level constraints and absolute-position constraints do. The real-model
runs add the harder lesson: a formalism can be expressive enough for the solver
and still too demanding for a small local model to produce cheaply and
faithfully.

The repair layer introduces another risk. A deterministic repair that uses the
source text is acceptable only if the transformation is specified before the
experiment, logged, and auditable. Otherwise, it could become an implicit
post-hoc reasoning channel. The current implementation logs repair attempts and
uses source-grounded transformations. The replication artifact now includes a
source-linked audit packet for the BBH-derived VFR condition, but this should
be read as audit infrastructure rather than as an independent blinded human
audit. A stronger study should add independent annotation of formalization
faithfulness before drawing broad conclusions about translation reliability.
The automatic traceability checks added here reduce, but do not remove, this
threat. They can detect missing entities, unsupported rules, ungrounded source
spans, and mismatch against the deterministic benchmark formalizer. They cannot
decide every natural-language paraphrase or pragmatic entailment that a human
annotator might judge acceptable.

The added automatic robustness checks should be read in the same spirit. The
BBH-heldout deterministic ablation shows that the finite-domain formalism is
not tuned only to the main 120 public-source instances, and the repeated local
probe gives a useful sanity check on Qwen stability. However, the repeated
probe remains modest relative to the full matrix and exposes a negative Gemma
pattern on the sampled instances. It strengthens the paper's boundary
conditions rather than supporting a broader model-agnostic claim.

The local-model runs are also not bit-for-bit reproducible. Deterministic data
generation, heldout selection, audit sampling, and bootstrap intervals are
seeded, but LM Studio decoding is controlled through prompts, temperatures,
token budgets, and model identifiers rather than a backend-level random seed.
This does not invalidate the reported aggregate comparisons, because the
experiments log all outputs and the repeated probe directly tests stability
under fresh local calls. It does limit the strength of any claim that an
independent rerun will reproduce every individual sampled answer exactly.

Finally, the resource accounting remains incomplete. The study reports model
calls, tokens, serial latency, and per-problem variability for process-level
RSS and runtime metadata. It does not measure energy, thermal throttling,
hardware counters, or controlled repeated-session peak memory under fixed
system load. Future claims about low-resource deployment should add these
measurements before making stronger hardware-level conclusions.

\paragraph{Threats to validity.}
The main validity threats are specific enough to be stated directly. First,
there is a construct-validity threat in the resource metrics: tokens, calls,
and serial latency are meaningful proxies for local decoding effort, but they
are not equivalent to energy or memory pressure. Second, there is an
internal-validity threat in the repair module: even deterministic repairs can
look like hidden reasoning if their preconditions are not inspectable. The
implementation mitigates this by using source spans, logging repair attempts,
and separating basic SLM--solver from VFR-LLM, but the threat is not eliminated.
Third, there is an external-validity threat in the benchmark design. The
BBH-derived subsets are public-source and useful, yet they are filtered to the
formal fragment evaluated here. Fourth, there is a conclusion-validity threat
in the model-family interaction: Qwen and Gemma do not behave the same way on
the extended public-source condition. For this reason, the difficulty analysis
is reported descriptively and the paper avoids a general scaling or
model-agnostic claim. Fifth, the baseline-strength concern is now addressed
empirically rather than left open. In addition to serial vanilla
self-consistency with five samples, we evaluate a cost-aware adaptive
self-consistency baseline that stops sampling early under the two-class Beta
criterion of \cite{aggarwal2023adaptive}. As reported in
Table~\ref{tab:adaptive-self-consistency}, this baseline reduces calls and
tokens at comparable accuracy for both Qwen and Gemma but does not close the
single-call accuracy gap with VFR-LLM on the precedence and Qwen BBH-extended
strata, while remaining statistically indistinguishable on exactly the strata
where the paper already declines to claim an advantage, including Gemma
BBH-extended. The residual threat is that even more
aggressive cost-aware schedules, or larger sample caps with difficulty-adaptive
budgets \cite{wang2025difficulty}, could further narrow the resource gap; a full
sweep over adaptive schedules is left to future work.

\section{Conclusion}

This study asks whether verifiable symbolic reasoning can reduce local LLM
inference cost compared with repeated-generation strategies. The answer from
the current evidence is not a simple yes or no. It is positive in several
bounded strata, marginal in one public-source Gemma condition, and blocked or
negative in others.

On 120 pure-precedence problems, VFR-LLM reaches 0.983 accuracy with one model
call, compared with 0.700 accuracy and five calls for serial self-consistency.
It also reduces total tokens by 34.4\% relative to self-consistency. This is
the strongest supported contribution: when direct answering is not accurate
enough and repeated sampling is the practical alternative, a traceable
SLM--solver pipeline can preserve or improve accuracy while reducing measured
local-compute proxies.

The public-data checks point in the same direction for Qwen. On 60 pairwise
instances curated from BIG-Bench Hard Logical Deduction, VFR-LLM reaches 0.967
accuracy, compared with 0.867 for self-consistency and 0.850 for direct
answering, while reducing total tokens by 49.2\% relative to self-consistency.
On the harder 120-instance BBH-derived extended subset, Qwen VFR-LLM reaches
0.933 accuracy versus 0.283 for self-consistency and 0.317 for direct
answering. This second public-source result is stronger in accuracy but more
nuanced in resources: VFR reduces calls and serial latency, but not total
tokens. These results do not show that the method solves full BBH logical
deduction. They show that the positive Qwen result survives public-source
benchmarks when the task is filtered to the finite-domain logic that the method
actually implements.

The model-family results set a clear boundary. On Gemma pure-precedence tasks,
VFR-LLM also improves accuracy and reduces calls and tokens relative to
self-consistency, although it is still more expensive than direct answering.
On Gemma typed tasks, VFR improves over direct answering and self-consistency
but only modestly over the basic solver, and its latency remains high. On
Gemma BBH-extended, the VFR gain over direct answering is only 0.025 and is not
statistically significant. The Phi typed run is negative: the model often fails
to produce valid or useful formalizations, and VFR-LLM performs worse than
direct answering. These results do not invalidate the broader research
direction, but they reject the broad version of the compute-reduction
hypothesis for the current typed formalism.

The automatic traceability checks sharpen this boundary. The strongest Qwen
result coincides with high source grounding and normalized constraint recall,
whereas the weaker Gemma BBH-extended result coincides with much lower fully
traceable formalization rates. The deterministic negative controls also show
that the symbolic layer rejects impossible formal states instead of returning
plausible but invalid orders.

The added non-manual robustness checks make this conclusion more defensible.
The pipeline ablation shows that traceability without repair is too
conservative and that the full Qwen gain depends on the combined VFR policy.
The heldout deterministic BBH-derived set preserves the need for the extended
finite-domain formalism. The expanded repeated local probe preserves the Qwen
pattern but also confirms that Gemma should remain a cautionary robustness
condition.

The evidence therefore supports a bounded conclusion: VFR-LLM is a
resource-aware and auditable alternative to serial self-consistency on narrow
explicit-constraint tasks, and a diagnostic platform for studying when richer
symbolic targets become too expensive for local SLMs. The next technical step
is not simply to add more expressive logic. It is to design a typed formal
target that is more compact, easier for small local models to emit, and still
strong enough to expose solver-relevant structure. The evidence suggests that
the right question is not whether logic helps local models in general, but
which bounded logical interface gives a particular local model the best
accuracy--resource trade-off on a clearly specified task class. Two comparisons
would most sharpen this question: an empirical baseline against cost-aware
self-consistency rather than only serial vanilla sampling, and a non-ordering
finite-domain task family that tests whether the trade-off survives beyond
total-order problems.

\appendix
\section{Prompting and Execution Protocol}
\label{app:prompting-protocol}

This appendix records the prompting protocol used in the local-model runs. The
purpose is to make the experimental comparison reproducible without requiring a
reader to infer the prompt design from the result tables.

\paragraph{Common request format.}
Each LM Studio call is sent to the local OpenAI-compatible chat endpoint with
\texttt{stream=false}. The request contains one system message and one user
message. Outputs are constrained with a JSON schema whenever the backend
accepts the OpenAI-compatible \texttt{response\_format} field. Answering
methods parse the response into \texttt{final\_answer} and
\texttt{reasoning}. Formalization methods parse the response into the
formalization schema described in the main text.

\paragraph{Direct-answer template.}
The direct-answer user message asks the model to solve the instance and return
the order from first to last:

\begin{quote}
\small
Solve the problem and give the order from first to last. Include the problem
identifier, the problem text, and use the exact answer format
\texttt{item1 > item2 > item3}.
\end{quote}

The system message requires valid JSON with fields \texttt{final\_answer} and
\texttt{reasoning}, keeps reasoning to one short sentence, and requires the
same order format. The decoding temperature is 0.0 and the answer budget is
128 generated tokens.

\paragraph{Chain-of-thought template.}
The chain-of-thought user message differs from direct answering only in the
instruction to reason before producing the final answer:

\begin{quote}
\small
Solve the problem step by step, then give the final order from first to last.
Include the problem identifier, the problem text, and use the exact final
answer format \texttt{item1 > item2 > item3}.
\end{quote}

The output schema, system message, temperature, and answer token budget are the
same as in direct answering. The recorded output is still the parsed final
answer; intermediate reasoning is not used by the solver.

\paragraph{Self-consistency template.}
Self-consistency repeats the chain-of-thought template \(k=5\) times with
temperature 0.7. Each sample is parsed through the same answer schema. The
predicted answer is the most common normalized final order among the five
samples. No solver or symbolic constraint check is applied to these samples.
The reported calls, tokens, and latency are the sum of the five serial samples.

\paragraph{Formalization template.}
The formalization user message asks the model to formalize the same problem and
includes the problem identifier, problem text, and expected domain when present:

\begin{quote}
\small
Formalize this precedence reasoning problem. Include the problem identifier,
the problem text, and the expected domain. Return a formalization with source
spans for every constraint.
\end{quote}

The system message requires valid JSON matching the formalization schema and
specifies the target language. Pairwise order is represented with
\texttt{before(A,B)} and \texttt{after(A,B)}. Generic type-level precedence is
represented with \texttt{type\_before(T1,T2)} plus entity types when the source
text states them. Absolute or ordinal positions are represented with
\texttt{position(A,N)}, using one-indexed positions in the requested
first-to-last order. Each constraint must include \texttt{source\_text},
\texttt{formal\_rule}, \texttt{kind}, and \texttt{confidence}. The prompt asks
for compact JSON and prefers type-level rules over enumerating all pairwise
constraints when the text licenses a type-level statement. The temperature is
0.0. The formalization token budget is 1024 by default and 1280 for longer
extended or Gemma conditions.

\paragraph{Basic solver versus VFR-LLM.}
The basic SLM--solver and VFR-LLM receive the same formalization prompt. The
basic solver applies schema validation and syntax normalization, then sends the
formalization to the solver. VFR-LLM additionally checks whether formal
constraints are grounded in the source text and applies only deterministic
local repairs: canonicalizing comparison syntax, removing unsupported
introductory rules, and correcting inverted binary arguments when the cited
source span explicitly licenses the correction. The repair module receives no
gold answer and does not call the model again.

\paragraph{Normalization and comparison.}
All final answers are normalized into a single delimiter convention before
scoring. Whitespace around \texttt{>} is ignored, but entity names are not
mapped through an answer-specific oracle. If the model emits an answer that
cannot be parsed into the expected order format, the instance is counted as an
answer or formalization failure according to the method that produced it.


\bibliographystyle{unsrtnat}
\bibliography{article/references}

\end{document}